\documentclass[twocolumn,aps,pra,superscriptaddress]{revtex4-1}
\usepackage{standalone}
\usepackage{nccfoots}
\usepackage{graphicx}
\usepackage{braket}
\usepackage{color}

\definecolor{mygreen}{rgb}{0,0.5,0}
\definecolor{myblue}{rgb}{0,0,0.75}
\definecolor{mymagenta}{cmyk}{0,1,0,0.12}
\usepackage[colorlinks=true,
   linkcolor=blue,
   citecolor=blue,
   urlcolor =blue]{hyperref}

\newcommand{\hc}{\widetilde{\text{H.c.}}}
\newcommand{\be}{\begin{equation}}
\newcommand{\ee}{\end{equation}}

\newcommand{\cf}{{\it cf.}}
\newcommand{\ie}{{\it i.e.}}

\expandafter\let\csname equation*\endcsname\relax

\expandafter\let\csname endequation*\endcsname\relax

\usepackage{amsmath}

\begin{document}
\title{Quantum State Transfer via Noisy Photonic and Phononic Waveguides}
\author{B. Vermersch}

\thanks{These two authors contributed equally.}

\affiliation{Institute for Theoretical Physics, University of Innsbruck, A-6020, Innsbruck, Austria}
\affiliation{Institute for Quantum Optics and Quantum Information of the Austrian Academy of Sciences, A-6020 Innsbruck, Austria}

\author{P.-O. Guimond}

\thanks{These two authors contributed equally.}

\affiliation{Institute for Theoretical Physics, University of Innsbruck, A-6020, Innsbruck, Austria}
\affiliation{Institute for Quantum Optics and Quantum Information of the Austrian Academy of Sciences, A-6020 Innsbruck, Austria}

\author{H. Pichler}
\affiliation{ITAMP, Harvard-Smithsonian Center for Astrophysics, Cambridge, Massachusetts 02138, USA}
\affiliation{Physics Department, Harvard University, Cambridge, Massachusetts 02138, USA}

\author{P. Zoller}
\affiliation{Institute for Theoretical Physics, University of Innsbruck, A-6020, Innsbruck,
Austria}
\affiliation{Institute for Quantum Optics and Quantum Information of the Austrian Academy
of Sciences, A-6020 Innsbruck, Austria}

\begin{abstract}  {We} describe a quantum state transfer protocol, where a quantum state of photons stored in a first cavity can be faithfully transferred to a second distant cavity via an infinite 1D waveguide, while being immune to arbitrary noise (e.g. thermal noise) injected into the waveguide. 
We extend the model and protocol to a cavity QED setup, where atomic ensembles, or single atoms representing quantum memory, are coupled to a cavity mode. We present a detailed study of sensitivity to imperfections, and apply a quantum error correction protocol to account for random losses (or additions) of photons in the waveguide. Our numerical analysis is enabled by matrix product state techniques to simulate the complete quantum circuit, which we generalize to include thermal input fields. Our discussion applies both to photonic and phononic quantum networks. 
\end{abstract}
\date{\today}
\maketitle

\textit{Introduction.---} 
The ability to transfer quantum states between
distant nodes of a quantum network via a quantum channel is a basic task in quantum information processing~\cite{Kimble2008,Northup2014,Hammerer2010,Reiserer2015}. An outstanding challenge is to achieve quantum state transfer~\cite{Cirac1997,nikolopoulos2013quantum} (QST) with high fidelity despite the presence of noise and decoherence in the quantum channel. In a quantum optical
setup the quantum channels are realized as 1D waveguides, where quantum information is carried by `flying
qubits' implemented either by photons in the optical~\cite{Ritter2012,Tiecke2014,Goban2015} or microwave regime~\cite{Eichler2012,VanLoo2013,Wenner2014,Grezes2014},
or phonons~\cite{Eisert2004,Hatanaka2014}. Imperfections in
the quantum channel thus include photon or phonon loss, and, in particular
for microwave photons and phonons, a (thermal) noise background~\cite{Habraken2012}. In
this Letter we propose a QST protocol and a corresponding quantum
optical setup which allow for state transfer with high fidelity,
undeterred by these imperfections. A key feature is that our protocol
and setup are \emph{a priori} immune to quantum or classical noise
\emph{injected} into the 1D waveguide, while imperfections such as
random generation and loss of photons or phonons during transmission
can be naturally corrected with an appropriate quantum error correction (QEC)
scheme~\cite{Michael2016}. 
\begin{figure}
\includegraphics[width=\columnwidth]{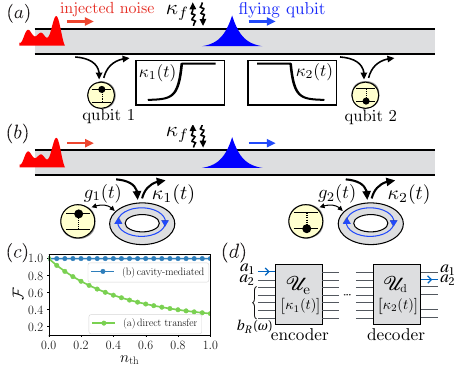}
\caption{{\it Quantum state transfer (QST) via a noisy waveguide.}\label{fig:setup}
(a)~QST where qubits are coupled {\em directly} with chiral coupling to a waveguide representing the quantum channel.
(b)~QST in a cavity QED setup, where atoms representing qubits are coupled to the waveguide with a cavity as mediator. 
(c)~Fidelity $\cal F$ for QST of a qubit as a function of photon occupation $n_\mathrm{th}$ representing a thermal noise injected into the waveguide for setups (a) and (b). For the protocol described in the text, setup (b) is robust to injected noise. (d)  `Write' of a quantum state from cavity~$1$ to a temporal mode in the (noisy) waveguide, and `Read' back to cavity~$2$ as a linear multimode encoder and decoder with encoding [decoding] functions $\kappa_1(t)$ [$\kappa_2(t)$] (see text).}
\end{figure}

The generic setup for QST in a quantum optical
network is illustrated in Fig.~\ref{fig:setup} as  transmission of a qubit state from a first to a second distant two-level atom via an infinite 1D bosonic open waveguide.
The scheme of Fig.~\ref{fig:setup}(a) assumes a \emph{chiral
coupling} of the two-level atoms to the waveguide~\cite{Lodahl2017,Bliokh2015}, as demonstrated in recent experiments with atoms~\cite{Mitsch2014} and quantum dots~\cite{Sollner2015}. The atomic
qubit is transferred in a decay process with a time-varying coupling to a \emph{right-moving} photonic (or phononic) wavepacket propagating in the waveguide, \ie~$\big(c_g\ket{g}_{1}+c_e\ket{e}_{1}\big)\ket{0}_{p}\rightarrow\ket{g}_{1}\big(c_g\ket{0}_{p}+c_e\ket{1}_{p}\big)$ where $\ket{}_1$ and $\ket{}_p$ denote the atomic and channel states. The transfer of the qubit state is then completed by reabsorbing
the photon (or phonon) in the second atom via the inverse operation, essentially mimicking the time-reversed process of the initial decay. Such transfer protocols have
been discussed in the theoretical literature~\cite{Cirac1997,Stannigel2010,nikolopoulos2013quantum,Ramos2016,Yao2013,Dlaska2017}, and demonstrated in recent experiments~\cite{Ritter2012}. A central assumption underlying these studies is, however, that the waveguide is initially prepared in the vacuum state, \ie~at zero temperature, and --~as shown in Fig.~\ref{fig:setup}(c)~-- the fidelity for QST (formally defined in App.~\ref{sec:avfid}) will degrade significantly in the presence of noise, e.g. thermal~\cite{Habraken2012}. Below we show that a simple variant of the setup with a cavity as mediator makes the QST protocol immune against arbitrary injected noise~\cite{Cirac1997,Clark2003} [\cf~Fig.~\ref{fig:setup}(b)].  Robust QST also provides the basis for distribution of entanglement in a quantum network. 

\textit{Photonic quantum network model.---} We consider the setup illustrated in  Fig.~\ref{fig:setup}(b), where each `node' consists of a two-level atom as qubit coupled to
a cavity mode. We assume that the cavity QED setup is designed with
a chiral  light-matter interface with coupling to  right-moving modes of the waveguide \footnote{The use of cavities also allows to efficiently decouple the atoms from unwanted emission into  non-guided modes~\cite{Sayrin2015}}\nocite{Sayrin2015}.
In the language of quantum optics the setup of Fig.~\ref{fig:setup}(b) is a cascaded quantum system~\cite{gardiner2015}, where the
first node is unidirectionally coupled to the second one. The dynamics is described by a quantum stochastic Schr\"odinger equation
(QSSE)~\cite{gardiner2015} for the
composite system of nodes and waveguide as  $i\frac{d}{dt}\ket{\Psi(t)}=H(t)\ket{\Psi(t)}$ ($\hbar=1$). The Hamiltonian is \mbox{$H(t)=\sum_{j=1,2}H_{n_{j}}(t)+V(t)$}
with \mbox{$H_{n_{j}}(t)=i g_j(t)\left(a_{j}^{\dagger}\sigma_{j}^{-}-\mathrm{H.c.}\!\right)$}
the Jaynes-Cummings Hamiltonian for node $j=1,2$ in the rotating wave approximation (RWA). Here $a_{j}$ are annihilation
operators for the cavity modes and $\sigma_j$'s are Pauli operators for
the two-level atoms with levels $\ket{g}_{j},\ket{e}_{j}$. We assume that the cavities are tuned to resonance with the two-level
atoms ($\omega_{c}=\omega_{eg}$), and the Hamiltonian is written in
the rotating frame. The coupling of the first and second cavity (located at
$x_{1}<x_2$)  to the right-moving modes of the channel  is described by the interaction
Hamiltonian
\begin{eqnarray}
\nonumber V(t)\!  &=&i\sum_{j=1,2}\sqrt{\tfrac{\kappa_{j}(t)}{2\pi}}\int_{{\cal B}}d\omega b_{\mathrm{R}}^{\dagger}(\omega)e^{i(\omega-\omega_{c})t-i\omega x_{j}/c}a_{j} -\text{H.c.}\\
&\equiv  i&\big(\sqrt{\kappa_{1}(t)}b_{\mathrm{R}}^{\dagger}(t)a_{1}\!+\!\sqrt{\kappa_{2}(t)}b_{\mathrm{R}}^{\dagger}(t\!-\!\tau)e^{i\phi}a_{2}-\text{H.c.}\big) \label{eq:defHint}
\end{eqnarray} in the RWA.
Here $b_{R}(\omega)$ denotes the annihilation operators of the continuum
of right-moving modes with frequency $\omega$ within a bandwidth $\mathcal B$ around the atomic transition frequency, $c$ is the velocity of light, and $\kappa_{1,2}(t)$
is a decay rate to the waveguide. In the second line
of Eq.~\eqref{eq:defHint} we have rewritten this interaction in terms of quantum
noise operators $b_{\mathrm{R}}(t)$ satisfying white noise commutation relations
$[b_{\mathrm{R}}(t),b_{\mathrm{R}}^\dagger(s)]=\delta(t-s)$. The parameter $\tau=d/c$,
with $d=x_{2}-x_{1}>0$, denotes the time delay of the propagation between
the two nodes, and $\phi=-\omega_c\tau$ is the propagation phase.
For a cascaded quantum system with purely unidirectional
couplings, $\tau$ and $\phi$ can always be absorbed in a redefinition
of the time and phase of the second node. Noise injected into the waveguide
is specified by the hierarchy of normally ordered correlation functions
of $b_{\mathrm{R}}(t)$. In particular the Fourier transform of the correlation function $\langle b_{\mathrm{R}}^{\dagger}(t)b_{\mathrm{R}}(s)\rangle$ provides
the spectrum of the incident noise $S(\omega)$, which for white (thermal) noise corresponds
to $\langle b_{\mathrm{R}}^{\dagger}(t)b_{\mathrm{R}}(s)\rangle=n_\text{th}\delta(t-s)$ with
occupation number $n_\text{th}$ and flat spectrum $S(\omega)=n_\text{th}$.

\textit{Quantum state transfer protocol.---} 
To illustrate immunity to injected noise in QST we consider first a minimal model of a pair of cavities coupled  to the waveguide. The quantum Langevin equations (QLEs) for the annihilation operators of the two cavity modes $a_{1,2}(t)$ in the Heisenberg picture read [\cf~App.~\ref{sec:model}]
\begin{eqnarray}
\label{eqda1}\frac{d{a}_{1}}{dt}&=& -\dfrac{1}{2}\kappa_{1}(t){a}_{1}(t)-\sqrt{\kappa_{1}(t)}b_{\mathrm{R}}(t),\\
\frac{d{a}_{2}}{dt}&=& -\frac{1}{2}\kappa_{2}(t){a}_{2}(t)-\sqrt{\kappa_{2}(t)}\big[b_{\mathrm{R}}(t)+\sqrt{\kappa_{1}(t)}{a}_{1}(t)\big].\nonumber
\end{eqnarray}
These equations describe the driving of the first cavity by an input noise field $b_{\mathrm{R}}(t)$ \footnote{$b_{\mathrm{R}}(t)$ is always written in the interaction picture and acts as a source term for the nodes in \eqref{eqda1}.},
while the second cavity is driven by both $b_{\mathrm{R}}(t)$
and the first cavity. We can always find
a family of coupling functions $\kappa_{1,2}(t)$, satisfying
the time-reversal condition $\kappa_{2}(t)=\kappa_{1}(-t)$ [see inset Fig.~\ref{fig:setup}(a)], which achieves a mapping  
\begin{eqnarray}
\label{eq:mapping}a_{1}(t_{i})\rightarrow-a_{2}(t_{f}),
\end{eqnarray}
\ie~the operator of the first cavity mode at initial time $t_{i}$ is mapped 
to the second cavity mode at final time $t_{f}$, with no admixture from $b_{\mathrm{R}}(t)$ [\cf~App.~\ref{sec:mapping}].
In other words, an arbitrary photon superposition state prepared initially in the first cavity can be faithfully transferred to the second distant cavity without being contaminated by incident
noise. This result holds without any assumption on the noise statistics. It is intrinsically related to the linearity of the
above QLEs, which allows
the effect of noise acting equally on both cavities to drop
out by quantum interference.
The setup can thus be combined with other elements of linear optics, such as beamsplitters [\cf~App.~\ref{sec:mapping}].

Robustness of QST to injected noise generalizes immediately to more complex systems representing effective `coupled harmonic oscillators'.  We can then add atomic ensembles of $N$ two-level atoms represented by atomic hyperfine states~\cite{Julsgaard2004,Reimann2015,Hammerer2010} to the first and second cavities ($j=1,2$). 
Spin-excitations in atomic ensembles~\cite{Colombe2007,Brennecke2007}, generated by the collective spin operator $S_j^+=\frac{1}{\sqrt{N}}\sum_{i=1}^N\sigma_{i,j}^+$ with $i$ the sum over atomic spin-operators of node $j$, are again harmonic for low densities. Moreover, they can be coupled in a Raman process to the cavity mode,  $H_{n_j}= \tilde g_j (t) (S_j^+ a_j +\mathrm{H.c.})$, as familiar from the read and write of photonic quantum states to atomic ensembles as quantum memory
\footnote{Adiabatic elimination of the cavity provides QLEs for $S^+_j$ analogous to (\ref{eqda1}).}. 
This provides a way of getting an effective time-dependent coupling to the waveguide in a setup with constant cavity decay.
Our protocol thus generalizes to transfer of quantum states stored as long-lived spin excitation in a first atomic ensemble to a second remote ensemble [\cf~App.~\ref{sec:atomicensembles}].

Returning to the setup of Fig.~\ref{fig:setup}(b) with a single atom as qubit coupled to a cavity mode, we achieve -- in contrast to the setup of Fig.~\ref{fig:setup}(a) -- QST immune to injected noise in a three step process. (i) We first map the atomic qubit state $c_g\ket{g}_{1}+c_e\ket{e}_{1}$ to the cavity mode $c_g\ket{0}_{1}+c_e\ket{1}_{1}$ with the cavity decoupled from the waveguide
 \footnote{The cavity modes can be prepared initially in a vacuum state via a dissipative optical pumping process with atoms, analogous to Ref.~\cite{Habraken2012}}. (ii) With atomic qubits decoupled from cavities we transfer the photon superposition state to the second cavity as above~\footnote{If there is an imperfection in step (i), the resulting mixed state is transferred without additional error to the second cavity.}. (iii)~We perform the time-inverse of step (i) in the second node. This QST protocol generalizes to several atoms as a quantum register representing an entangled state of qubits, which can either be transferred sequentially, or mapped collectively to a multiphoton superposition state in the cavity, to be transferred to the second node~\footnote{This is achieved, for example, with quantum logic operations available with trapped ions stored in a cavity~\cite{Casabone2015}.}\nocite{Casabone2015}.
As depicted in Fig.~\ref{fig:setup}(d), we can understand our QST protocol in the chiral cavity setup [Fig.~\ref{fig:setup}(b)], consisting of a {\em  write operation} of the qubit in the first cavity to the waveguide as a quantum data bus, followed by a {\em read} into the second cavity. This `write' and `read' are both linear operations on the set of operators consisting of cavity and waveguide modes, or as an encoder and decoder into temporal modes specified by $\kappa_{1,2}(t)$, and physically implemented by the chiral cavity-waveguide interface.

\textit{Numerical techniques.---} We now study the sensitivity of the above protocol to errors. Imperfections may arise from inexact external control parameters including timing and deviations from perfect chirality. Moreover, loss or addition of photons can occur during propagation. We describe below a 
QEC scheme which corrects for such single photon errors.

\begin{figure}
\includegraphics[width=\columnwidth]{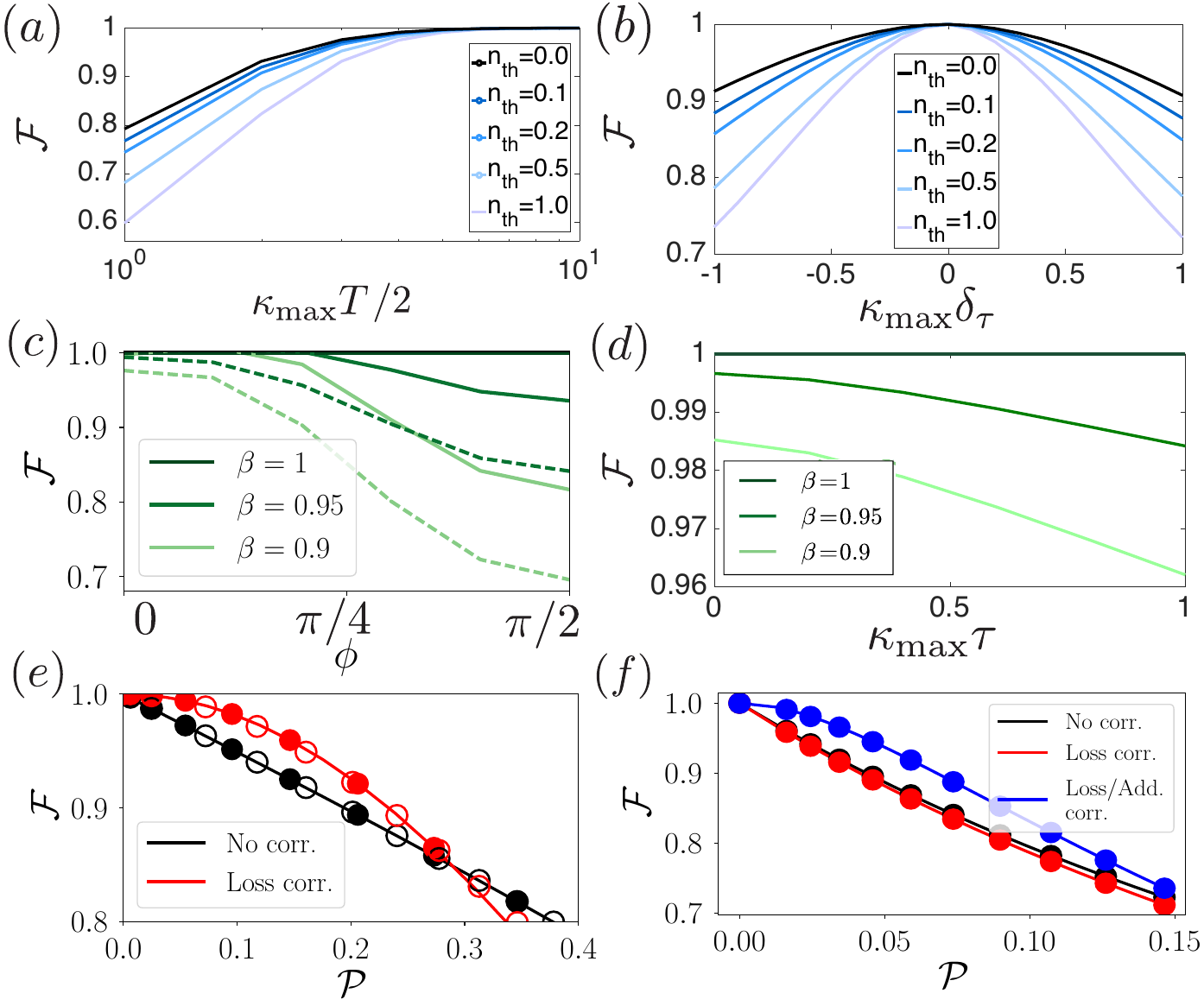}
\caption{{\it Role of imperfections.}  (a) Effect on the fidelity of a finite transfer time $T=t_f-t_i$, and (b) of an imperfect timing of $\kappa_2(t)$. 
(c) Fidelity as a function of $\phi$ for different $\beta$ factors and $\kappa_{\max}\tau\approx0$ (see text). Solid lines: $n_\text{th}=0$. Dashed lines: $n_\text{th}=0.25$. The fidelity is maximal when $\phi$ is multiple of $\pi$.
(d) By increasing $\kappa_{\max}\tau$ for $\phi=0$ the fidelity decreases.
(e)  QEC with the setup subject to waveguide and cavity losses. (f) QEC with the setup coupled to a reservoir with photon occupation $n_\text{th}'=1$. Black: no error correction. Red: correction against single photon losses. Blue: correction against single photon losses or additions. Solid lines: $n_\text{th}=0$, $\kappa'=0$. Full circles: $n_\text{th}=0.5$, $\kappa'=0$. Empty circles: $n_\text{th}=0$, $\kappa_f=0$. \label{fig:error}} 
\end{figure}

A study of imperfections in QST will necessarily be numerical in nature, as it requires solution of the QSSE with injected noise accounting for nonlinearities in atom-light coupling. 
Beyond Eq.~\eqref{eq:defHint}, the Hamiltonian must include coupling to both right- and left-propagating modes in the waveguide, and should account for possible couplings of waveguide and cavities to additional reservoirs representing decoherence [\cf~App.~\ref{sec:model}].  We have developed and employed three techniques to simulate the complete dynamics of the quantum circuits as depicted in Figs.~\ref{fig:setup}(a) and \ref{fig:setup}(b). First, we use matrix product states (MPS) techniques to integrate the QSSE discretized in time steps, as developed in Ref.~\cite{Pichler2016}, which we generalize to include injected quantum noise.  Our method allows a general input field to be simulated using purification techniques, by entangling time-bins of the photonic field with ancilla copies in the initial state (for related techniques developed in condensed matter physics see Ref.~\cite{Schollwock2011}).
This method also allows the study of non-Markovian effects (\ie~for finite retardation $\tau>0$) in the case of imperfect chiral couplings, and is well suited to represent various kinds of noise.
Second, we solve the master equation describing the nodes, which allows for efficient simulations valid in the Markovian limit.
Finally, to simulate the QST in non-chiral setups as described at the end of this Letter, we solve the dynamics of the nodes and of a discrete set of waveguide modes, following Ref.~\cite{Ramos2016}. For a detailed description of the complete model and numerical methods we refer to Apps.~\ref{sec:model} and \ref{sec:MPS}, and present below our main results assuming thermal injected noise $n_{\rm th}$. 

{\it Sensitivity to coupling functions $\kappa_{1,2}(t)$.---} 
In Figs.~\ref{fig:error}(a) and \ref{fig:error}(b) we study the sensitivity of QST to the functions $\kappa_{1,2}(t)$ for the minimal model of nodes represented by cavities. Figure~\ref{fig:error}(a) shows the effect of the protocol duration $T=t_f-t_i$ which in the ideal case is required to fulfill \mbox{$T\gg 1/\kappa_{\max}$}, with $\kappa_{\max}$ the maximum value of $\kappa_{1,2}(t)$. For finite durations, the effect on the fidelity scales linearly with the noise intensity but quickly vanishes for $\kappa_{\max}T \gtrsim 10$, above which $\mathcal F \geq 0.99$. In all other figures of this work we use $\kappa_{\max}T=20$.
In Fig.~\ref{fig:error}(b), we show the effect of an imperfection $\delta_\tau$ in the timing of the coupling functions, namely $\kappa_2(t)=\kappa_1(\delta_\tau-t)$. The digression from unity is quadratic in $\delta_\tau$ but linear in noise intensity. This result illustrates that only the proper decoding function allows one to unravel the quantum state emitted by the first cavity on top of the injected noise.
Note that in addition to errors in the coupling functions, the fidelity is also sensitive to the frequency matching of the cavities~\cite{Korotkov2011}, which we discuss in App.~\ref{sec:mismatch}.

{\it Imperfect chirality.---} For an optical fiber with chirally coupled resonators~\cite{Sayrin2015}, the nodes emit only a fraction $\beta<1$ of their excitations in the right direction.
The dynamics then also depends on the propagation phase $\phi$~\cite{Lodahl2017} and on the time delay $\tau$. As illustrated in Fig.~\ref{fig:error}(c), the effect of imperfect chirality in the Markovian regime ($\kappa_{\max}\tau\approx0$) crucially depends on $\phi$, as a consequence of interferences between the photon emissions of the two cavities in the left direction. In particular, for $\phi=0$, they interfere destructively, leading to a higher fidelity. This interference decreases for finite values of $\kappa_{\max}\tau$, as shown in Fig.~\ref{fig:error}(d).

{\it Quantum error correction.---}
In contrast to `injected' noise,  loss and injection of photons occurring during propagation between the two cavities represent decoherence mechanisms, which affect the fidelity of the protocol~\cite{Northup2014}.
Such errors can be corrected in the framework of QEC. Instead of encoding the qubits in the Fock states $\ket{0}$ and $\ket{1}$, we use multiphoton states, with the requirement that the loss or addition of a photon projects them onto a new subspace where the error can be detected and corrected. 
A possibility is to use a basis of cat states, \ie~superposition of coherent states~\cite{Haroche2006,Ourjoumtsev2006}, where a photon loss only induces a change of parity of the photon number~\cite{Ofek2016}. While we present the efficiency of  QST with cat states in App.~\ref{sec:qec}, we use here a basis of orthogonal photonic states for the qubit encoding~\cite{Michael2016}.

We first consider a protocol protecting against single photon losses. Here, the state of the first qubit is mapped to the first cavity as \mbox{$c_g \ket{g}_1 + c_e \ket{e}_1 \to c_g \ket{+}_1 + c_e \ket{-}_1$}, where the cavity logical basis \mbox{$\ket{\pm}_j=(\ket{0}_j\pm\sqrt{2}\ket{2}_j+\ket{4}_j )/2$} has even photon parity. This unitary transformation can be realized with optimal control pulses driving the qubit and the cavity while using the dispersive shift between the qubit and the cavity mode as nonlinear element~\cite{Ofek2016}. 
Waveguide losses, with rates $\kappa_f$, can be modeled with a beamsplitter with transmission probability $\exp(-\kappa_f \tau)$, whereas the rate of cavity losses is denoted $\kappa'$. The single photon loss probability is then $\mathcal P = 1-\exp(-\kappa_f \tau -\kappa' T)$. The density matrix $\rho_f$ of the second cavity at the end of the protocol reads
\begin{equation}
\rho_f = \ket{\Psi_0} \bra{\Psi_0} + \mathcal P \ket{\Psi_{-1}} \bra{\Psi_{-1}} + \mathcal{O}(\mathcal{P}^2),
\end{equation}
where the unnormalized states $\ket{\Psi_0}$ and $\ket{\Psi_{-1}}$, written explicitly in App.~\ref{sec:qec}, have even or odd parity, respectively, and satisfy $\ket{\Psi_{-1}}= a_2 \ket{\Psi_0}$.
The state $\ket{\Psi_0}$ corresponds to the case where no stochastic photon loss occurred, whereas the state $\ket{\Psi_{-1}}$ is obtained if one photon was lost in the process.
The last step of the protocol consists in measuring the photon number parity in the second cavity, and -- conditional on the outcome -- apply unitary operations transferring the photon state to qubit $2$.
As shown in Fig.~\ref{fig:error}(e), this encoding improves significantly the fidelity for small losses $\mathcal P \ll 1$, up to a threshold value $\mathcal P\approx 0.29$. Note that both protocols are insensitive to injected noise. 

We consider now a situation where the waveguide is coupled to a finite temperature reservoir with $n'_\text{th}=1$ thermal occupation number which stochastically adds and absorbs photons.
Here the qubit state is encoded as $c_g\ket{+}_1+c_e\ket{-}_1$, where $\ket{\pm}_1=(\ket{0}_1\pm\sqrt{2}\ket{3}_1+\ket{6}_1)/2$ have photon number $0$ modulo $3$. The state $\rho_f$ after the transfer is a mixture of $\ket{\Psi_k}\bra{\Psi_k}$ with $k=-1,0,+1$ corresponding to the cases of a single photon loss, of no photon loss or addition, and of a single photon addition. These states satisfy $\ket{\Psi_{-1}}=a_2\ket{\Psi_0}$ and $\ket{\Psi_{+1}}=a_2^\dagger\ket{\Psi_0}$ and are distinguishable by measurement of the photon number modulo $3$.
In the limit of small error probabilities, one retrieves the original qubit state by applying a unitary operation conditioned on the measurement outcome. In Fig.~\ref{fig:error}(f) we show that this protocol corrects the errors for $\mathcal P\ll 1$ independently of injected noise intensity. This approach extends to arbitrary number of photon losses and additions, although at the cost of a lower range of achievable $\mathcal P$~\cite{Michael2016}.

{\it Closed systems.---} 
\begin{figure}
\includegraphics[width=\columnwidth]{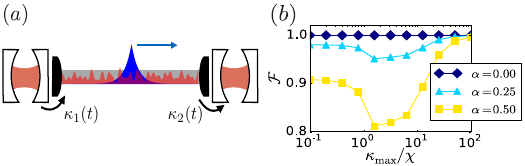}
\caption{{\it QST in non-cascaded systems.}
(a) QST in a closed system with two cavities coupled to a finite waveguide. (b) Fidelity as a function of the cavity nonlinearities $\chi$ and for different initial occupation of the waveguide. The fidelity approaches unity in the linear limit $\chi\to0$. 
\label{fig:chirality}}
\end{figure}
Our results can also be observed in closed systems  [\cf~Fig.~\ref{fig:chirality}(a)], where two cavities are coupled, for instance, via a finite optical fiber or a microwave transmission line \cite{Blais2004}. Note that in circuit QED setups, time-dependent couplings $\kappa_j(t)$ can be realized via tunable couplers~\cite{Korotkov2011,Wenner2014,Srinivasan2014}.
This system is not chiral, as the dynamics of the first cavity can be perturbed by reflections from the second one.
We numerically demonstrate robustness against noise, which is here represented as initial occupation of the waveguide. In addition, we consider the effect of Kerr nonlinearities, \ie~we add terms $-\chi a_j^\dagger a_j^\dagger a_j a_j$ [\cf~App.~\ref{sec:closed}],  which are relevant for circuit QED setups~\cite{Ofek2016}, to the Hamiltonian.
The results are presented in Fig.~\ref{fig:chirality}(b) with each (discrete) waveguide mode initially in a coherent state $\ket{\alpha}$.
QST becomes robust against noise in the transition from the cavity as an effective two-level system ($\chi\to\infty$) to perfect harmonic oscillator ($\chi\to0$).

{\it Conclusion.---} 
Robustness to arbitrary injected noise in transferring a quantum state between two cavities relies on the linearity of the write and read into temporal modes~[\cf~Fig.~\ref{fig:setup}(d)], with quantum noise canceled by quantum interference. While here we have focused on QST between two distant cavity modes, our approach generalizes to a setup involving many nonlocal bosonic resonator modes, which can be realized with various physical platforms, and as hybrid systems. 

{\it Note added---} A related setup and protocol have been proposed in an independent work by Z. L. Xiang {\it et~al.}~\cite{Xiang2016}.

\begin{acknowledgments}
BV and POG contributed equally to this work. HP provided the matrix product state code to integrate the QSSE, which was extended by POG to noisy input. We thank C.~Muschik, M.~Leib, K.~M\o lmer, B.~Vogell and G.~Kirchmair for  discussions. Simulations of the master equation were performed with the QuTiP toolbox~\cite{Johansson20131234}. Work at Innsbruck is supported by the EU  projects UQUAM and RYSQ, the SFB FOQUS of the Austrian Science Fund (FWF Project No. F4016-N23) and the Army Research Laboratory Center for Distributed Quantum Information via the project Scalable Ion-Trap Quantum Network (SciNet).
\end{acknowledgments}

\appendix

\section{Calculation of average fidelities for quantum state transfer}
\label{sec:avfid}
Throughout this work the fidelity of quantum state transfer (QST)  $\cal F$ is defined as the overlap between the final state of the second node at the end of the protocol and the state obtained in an ideal transfer, averaged over all initial qubit states for the first node. Formally, one can access this value by making use of a Choi-Jamiolkowski isomorphism between quantum processes and quantum states~\cite{Nielsen2002}, where the first node is prepared initially in an entangled state with an ancilla copy of itself which is otherwise completely decoupled from the dynamics. 

For the situation depicted in Fig.~1(b) of the main text, we focus on the transfer between the cavities via noisy waveguide.
The initial state of the first node thus reads
\begin{equation}
\ket{\psi(t_i)}_{1,a} = \big(\ket{0}_1\ket{0}_a+\ket{1}_1\ket{1}_a\big)/\sqrt{2},\label{eq:inistate}
\end{equation}
 where the index $1$ denotes the cavity and $a$ the ancilla, while the atom is decoupled.
The average fidelity is then obtained numerically by simulating the QST from cavity $1$ to cavity $2$, including imperfections. We denote the density matrix of second cavity and ancilla at the end of the protocol $\rho_{2,a}$ and its ideal value $\rho_\text{ideal}=\ket{\psi_\text{ideal}}_{2,a}\bra{\psi_\text{ideal}}$ with $\ket{\psi_\text{ideal}}_{2,a}=\ket{\psi(t_i)}_{2,a},$ up to propagation phase factors for the $\ket{1}_2\ket{1}_a$ component.
The fidelity then reads
\begin{equation}
\mathcal{F} = {Tr}\left(\sqrt{\sqrt{\rho_{a,2}}  \rho_\text{ideal} \sqrt{\rho_{a,2}}} \right)^2.
\end{equation}

For nodes realized with atomic ensembles of $N$ atoms as two-level systems, we apply a similar procedure, where the Fock states $\ket{0}_1$ and $\ket{1}_1$ in Eq.~\eqref{eq:inistate} are replaced with the collective ground state of the ensemble $\ket{G}_1$ and the excited state $\ket{E}_1=\frac{1}{\sqrt{N}}\sum_{i=1}^N \sigma_{i,1}^+ \ket{G}_1$, with $\sigma_{i,j}^+$ the creation operator for excitation of atom $i$ in the node $j$. In the quantum error correction (QEC) protocols, these states are replaced with the multiphoton states $\ket{\pm}_1$ as defined in the main text.

\section{Dynamics of cavities chirally coupled to a waveguide}\label{sec:model}

Here we provide details on the model of two cavities coupled to a waveguide with time-dependent decay rates, including imperfections such as deviation from unidirectionality, propagation losses and cavity decays. We present the corresponding form of the quantum stochastic Schr\"odinger equation (QSSE), the quantum Langevin equations (QLEs)  and the master equations supporting the analytical and numerical study of the QST dynamics.

\subsection{Model}

Our model consists of cavities as harmonic oscillators coupled on resonance to atomic two-level systems as qubits, and with chiral coupling to a waveguide. The dynamics is described by the Hamiltonian $H(t)=\sum_{j=1,2} H_{n_j}(t)+V_\beta(t)$, with the node Hamiltonians $H_{n_j}(t)$ as given in the main text,  and couplings to the waveguide:
\begin{eqnarray}
\nonumber V_\beta(t)  &=&i\sum_{j=1,2}\sqrt{\tfrac{\beta\kappa_{j}(t)}{2\pi}}\int_{{\cal B}}d\omega b_{{R}}^{\dagger}(\omega)e^{i(\omega-\omega_{c})t-i\omega x_{j}/c}a_{j} \\
&+& i \sum_{j=1,2}\sqrt{\tfrac{(1-\beta)\kappa_{j}(t)}{2\pi}} \int_{{\cal B}}d\omega b_{{L}}^{\dagger}(\omega)e^{i(\omega-\omega_{c})t+i\omega x_{j}/c}a_{j} \nonumber \\
&+&\text{H.c}.\label{eq:V} 
\end{eqnarray}
The Hamiltonian $H(t)$ is written in a frame where node operators $a_j$ and $\sigma_j^-$ rotate with the cavity frequency $\omega_c$, and in an interaction picture with the waveguide bare Hamiltonian $H_B=\int_\mathcal{B} d\omega\, \omega[b^\dagger_L(\omega)b_L(\omega)+b^\dagger_R(\omega)b_R(\omega)]$. Here $b_L(\omega)$ and $b_R(\omega)$ are left- and right-moving waveguide modes satisfying $[b_i(\omega),b^\dagger_{i'}(\omega')]=\delta_{i,i'}\delta(\omega-\omega')$, and we assumed a linear dispersion relation around $\omega_c$.
Moreover, in writing Eq.~\eqref{eq:V}, we have assumed under the Born-Markov approximation that the decay rates $\kappa_j(t)$ can be considered constant over the bandwidth $\mathcal{B}$. 
Here,  the parameter $0\leq\beta\leq1$ defines the chirality of the coupling with rates $\beta \kappa_j(t)$, respectively $(1-\beta)\kappa_j(t)$, to the right- and left-moving modes. For the perfectly chiral case ($\beta=1$), $V_\beta(t)$ correspond to the interaction Hamiltonian $V(t)$ as written in Eq.~(1) of the main text.

The interaction Hamiltonian can be rewritten as 
\begin{eqnarray}
V_\beta(t)&=& i\sum_{j=1,2}\sqrt{\beta\kappa_{j}(t)} b_{{R}}^{\dagger}\big(t-(j-1)\tau\big) e^{i(j-1)\phi} a_{j} \nonumber \\
&+& i\sum_{j=1,2} \sqrt{(1-\beta)\kappa_{j}(t)} b_{{L}}^{\dagger}\big(t+(j-1)\tau\big) e^{-i(j-1)\phi}  a_{j}  \nonumber \\
&+&\text{H.c}, 
\end{eqnarray}
where we have set $x_1=0$, $x_2=d$. Here $\phi=-\omega_c \tau$ is the propagation phase and $\tau=d/c$ is the time delay. The quantum noise operators
\begin{equation}
b_{R,L}(t) = \frac{1}{\sqrt{2\pi}} \int_{{\cal B}}d\omega\, b_{{R,L}}(\omega)e^{-i(\omega-\omega_{c})t}, \label{eq:b}
\end{equation}
satisfy $[b_{i}(t),b^\dagger_{j}(s)] = \delta_{i,j}\delta(t-s)$ and represent the incoming or `injected' light field interacting with the cavities. Finally, we include in our description the effect of cavity and propagation losses by adding to the interaction Hamiltonian $V_\beta$ coupling terms to additional channels~\cite{gardiner2004quantum}.

\subsection{Quantum stochastic Schr\"odinger equation}
The quantum stochastic Schr\"odinger equation (QSSE) describes the stochastic evolution of the system as
\begin{equation}
i\frac{d\ket{\Psi(t)}}{dt}=H(t) \ket{\Psi(t)}, 
\end{equation}
where $\ket{\Psi(t)}$ represents the wavefunction of the system of nodes and quantum channels, which can be interpreted within the framework of Ito or Stratonovich calculus~\cite{gardiner2015}.
As detailed in App.~\ref{sec:MPS}, the numerical simulation of QSSE can be performed using a matrix product state (MPS) ansatz for the wavefunction $\ket{\Psi(t)}$, where the state of the waveguide (e.g. vacuum, thermal or coherent) can be efficiently represented. 
It is particularly well-suited in the non-cascaded case $\beta<1$ and for long delay times ($\kappa_j \tau \gg 1$) where non-Markovian effects arise~\cite{Pichler2015,Ramos2016}.

\subsection{Quantum Langevin Equations}
\label{sec:langevin}

The quantum Langevin equations (QLEs), describing the dynamics of an arbitrary operator $a$ acting on the nodes in the Heisenberg picture, is the starting point of our analytical study of the QST dynamics. In the following, we present their explicit form for the various situations addressed in the main text.

\subsubsection{Ideal setup}

In the ideal setup the system has no (additional) losses and the coupling between cavities and waveguide is perfectly chiral with the two nodes decoupled from left-moving modes. In the Heisenberg picture the dynamics of an arbitrary operator $a$ acting on the nodes can be obtained by formal integration of the Heisenberg equation for right-moving modes $b_R(\omega)$ around the resonant frequency $\omega_c$ (see for instance Ref.~\cite{gardiner2015}), which leads to
\begin{align}
\dot a =
&-\sum_{j=1,2} [a, a_j^\dagger] \left(\frac{\kappa_j(t)}{2}a_j+\sqrt{\kappa_j(t)} b_{R} (t) + g_j(t) \sigma_j^- \right) \nonumber\\
&-\sum_{j=1,2}[a,\sigma_j^+]  g_j(t) a_j- [a, a_2^\dagger] \sqrt{\kappa_2(t)\kappa_1(t)}  a_1  +\hc,
 \label{eq:langevin}
\end{align}
where we redefined the time of the second cavity to eliminate the time delay and absorbed the propagation phase in the definition of the cavity operators $a_j$.
Here, the notation $\hc$ refers to taking the complex conjugate of the total expression, except for operator $a$. Note that the operator $b_R(t)$ is expressed as in Eq.~\eqref{eq:b}, \ie~in the interaction picture, and represents here a source term driving the cavities. 
For $a=a_1$ and $a=a_2$, we obtain Eqs.~(2) of the main text when atoms and cavities are decoupled [$g_j(t)=0$].

\subsubsection{Imperfect chirality}
We now consider the case where the chiral coupling between nodes and waveguide is not perfect (\ie~unidirectional), that is the nodes also couple to left-moving modes $b_{{L}}(t)$.
In order to obtain a system of coupled local differential equations, we neglect the time delay $\tau$ assuming the Born-Markov approximation $\kappa_{\max}\tau \ll 1$, with $\kappa_{\max}$ the maximum value of $\kappa_1(t)$. The QLEs then read~\cite{Pichler2015}
\begin{eqnarray}
\dot a &=& -\sum_{j=1,2} [a, a_j^\dagger] \left(\frac{\beta\kappa_j(t)}{2}a_j+\sqrt{\beta\kappa_j(t)} b_{R} (t) \right) \nonumber \\
&-& [a, a_2^\dagger] \left(\beta\sqrt{\kappa_2(t) \kappa_1(t)}  a_1\right) , \nonumber \\
&-& \sum_{j=1,2} [a, a_j^\dagger] \left(\frac{(1-\beta)\kappa_j(t)}{2}a_j \right.  \label{eq:langevinlchiral}\\
&&\hspace{1cm}\left. +\sqrt{(1-\beta)\kappa_j(t)} e^{2i\phi\delta_{j,2}}  b_{L} (t) \right)  \nonumber \\
&-& [a, a_1^\dagger] \left(e^{-2i\phi}(1-\beta)\sqrt{\kappa_1(t) \kappa_2(t)}  a_2\right) +\hc \nonumber,
\end{eqnarray} 
where we redefined $b_L(t)\to e^{-i\phi} b_L(t-\tau)$ and assumed $g_j(t)=0$.
Note the importance of the propagation phase $\phi$ in the above equation, which governs the interference between the emission of the two cavities.

\subsubsection{Additional losses}

In the case of cavity and waveguide losses, the QLEs can be written as~\cite{gardiner2004quantum}
\begin{eqnarray}
\dot a &=& - [a, a_1^\dagger] \left(\frac{\kappa_1(t)}{2}a_1+\sqrt{\kappa_1(t)}  b_{R} (t) \right)  \nonumber \\
&-&  [a, a_2^\dagger] \left(\frac{\kappa_2(t)}{2}a_2+\sqrt{\kappa_2(t)} \cos(\theta) b_{R} (t) \right)  \nonumber \\
&-& [a, a_2^\dagger] \left(\cos(\theta)\sqrt{\kappa_2(t) \kappa_1(t)}  a_1\right) , \nonumber \\
&-& [a, a_2^\dagger]  \left( \sqrt{\kappa_2(t)}\sin(\theta) b_{U}(t) \right) \nonumber \\
 &-&\sum_{j=1,2} [a, a_j^\dagger] \left(\frac{\kappa'}{2}a_j+\sqrt{\kappa'} b_j'(t) \right)+\hc \label{qe:lamgevinlosses},
\end{eqnarray}
with $g_j(t)=0$, and where we have assumed perfect chirality ($\beta=1$). Here $\kappa'$ is the coupling of each cavity to unwanted (non-guided) modes with input fields $b_{1,2}'(t)$, and the waveguide losses with rates $\kappa_f$ are modelled by a beamsplitter mixing to an additional waveguide with $\cos(\theta)=\exp(-\kappa_f\tau/2)$ and input field $b_{U}(t)$.

\subsection{Master equation}

The master equation allows to perform numerical simulations of the evolution of the node reduced density matrix $\rho(t)={Tr}_\text{B}\big(\ket{\Psi(t)}\bra{\Psi(t)}\big)$, where $\text{Tr}_\text{B}$ denotes the trace over propagating modes. The mapping from the QLEs to the master equation can be realized using different techniques~\cite{gardiner2004quantum,vogel2006,gardiner2015}. To do so, one writes the QLE in the following form
\begin{eqnarray}
\dot  a &=&- \sum_\alpha [a,c_\alpha^\dagger]d_\alpha - \sum_k [a,e_k^\dagger] b_k(t) + \hc ,\label{eq:langevincanonical}
\end{eqnarray}
where the $\alpha$ index refers to interactions involving node operators and the $k$ index to input fields $b_k(t)$.
For instance, in the case of Eq.~\eqref{eq:langevin}, we have $c_1=a_1$, $d_1=(\kappa_1/2) a_1$,  $c_2=a_2$, $d_2=(\kappa_2/2) a_2$, $c_3=a_2$, $d_3=\sqrt{\kappa_1\kappa_2}a_1$, \mbox{$e_1=\sqrt{\kappa_1}a_1+\sqrt{\kappa_2}a_2$}, $b_1=b_{R}(t)$.

In the case where each channel is in a thermal state with $n_{th}(k)$ average photons, we obtain the master equation~\cite{gardiner2004quantum,vogel2006,gardiner2015}
\be
\frac{d}{dt} \rho = \sum_\alpha \mathcal{L}_\alpha(\rho) +\sum_k \mathcal{L}_k(\rho),
\ee where
\begin{eqnarray}
\mathcal{L}_\alpha(\rho) =&& [d_\alpha \rho, c_\alpha^\dagger]+[c_\alpha,\rho d_\alpha^\dagger], \\
\mathcal{L}_k(\rho) =&& n_{th}(k) \left( \mathcal{D}[e_k](\rho)+\mathcal{D}[e_k^\dagger](\rho) \right),
\end{eqnarray}
with $\mathcal{D}[e](\rho)=e \rho e^\dagger-\tfrac12(e^\dagger e \rho+\rho e^\dagger e)$.

\section{Operator mapping $a_{1}(t_{i})\rightarrow-a_{2}(t_{f})$  }\label{sec:mapping}

Here we show that the QST protocol between two nodes as linear harmonic oscillators performs the operator mapping of Eq.~(3) of the main text, assuming the ideal scenario where couplings to the environment are negligible ($\kappa'=\kappa_f=0$) and couplings to the waveguide are perfectly chiral ($\beta=1$). The system is described by Eqs.~(2) of the main text.

\subsection{Integration of the quantum Langevin equations}
In the input-output formalism, the output field of the first node in the waveguide is given by $b_{R}(t)+v(t)$, with the contribution from the first node $v(t)=\sqrt{\kappa_1(t)}a_1(t)$.
This variable represents the emission of the first node, containing the information about its initial quantum state. 
Conversely, the output field of the two cavities is given by $b_{R}(t)+w(t)$, where the contribution of the two nodes is $w(t)=v(t)+\sqrt{\kappa_2(t)}a_2(t)$. 
Using the Langevin equation~\eqref{eq:langevin}, these two variables evolve according to
\begin{eqnarray}\label{eq: motionv}
\frac{dv}{dt}&=&f_1 v-\kappa_1 b_{R}(t) \label{eq:motionv} \\ 
\frac{dw}{dt}&=&f_2 w +(f_1-\kappa_2-f_2)v-(\kappa_1+\kappa_2) b_{R}(t) \label{eq:motionw}
\end{eqnarray}
where $f_j=(\dot \kappa_j/\kappa_j-\kappa_j)/2$.
We require that these functions satisfy the condition \be \label{eq:req1}f_1(t)=f_2(t)+\kappa_2(t),\ee so that the equations for $v$ and $w$ decouple. 
In the standard QST description, \ie~without injected noise, this condition implies that the output field $w$ vanishes, or, in other words, that the second cavity absorbs all the emission of the first one.

The general solution of Eqs.~\eqref{eq: motionv},\eqref{eq:motionw} reads
\begin{eqnarray}
 \label{eqsolv}
v(t) &= &v(t_i)e^{\int_{t_i}^tdt' f_1(t')}\\ &&-\int_{t_i}^{t}dt'\, e^{{\int_{t'}^{t}dt'' f_1(t'')}} \kappa_1(t')  b_{R}(t'),\nonumber  \\
w(t) &=& w(t_i)\, e^{\int_{t_i}^tdt' f_2(t')} \\ &&-\int_{t_i}^{t}dt'\, e^{{\int_{t'}^{t}dt'' f_2(t'')}} \big(\kappa_1(t')+\kappa_2(t')\big) b_{R}(t') \nonumber. 
\end{eqnarray}
The second node operator $a_2(t)=(w(t)-v(t))/\sqrt{\kappa_2(t)}$ can then be written in the form
\be\begin{aligned} \label{eqSM:a2t}a_2(t_f)=& G_1(t_f,t_i) a_1(t_i) + G_2(t_f,t_i)a_2(t_i)\\&+\int_{t_i}^{t_f}dt'\, G(t,t') b_{R}(t'),\end{aligned}\ee
 where the node propagators are expressed as
\begin{eqnarray}
 \label{G12}
G_1(t,t') &= & \sqrt{\frac{\kappa_1(t')}{\kappa_2(t)}} e^{\int_{t'}^tdt'' f_1(t'')}\left(e^{-\int_{t'}^tdt'' \kappa_2(t'')}-1\right) \nonumber \\ 
G_2(t,t') &= & e^{-\frac12\int_{t'}^tdt'' \kappa_2(t'')},
\end{eqnarray}
and the noise propagator as
\begin{eqnarray}
 \label{eq:smdefG}
G(t,t') &= & -\sqrt{\kappa_1(t')} G_1(t,t') - \sqrt{\kappa_2(t')} G_2(t,t') \label{eq:G}.
\end{eqnarray}
In the limit of transfer times $T=t_f-t_i$ large compared to the typical emission rates $\sim 1/\kappa_{\max}$, the node propagators satisfy (see for example Ref.~\cite{Stannigel2011})
\begin{eqnarray}
 G_1(t_f,t_i)&=&-1\\G_2(t_f,t_i)&=&0,
\end{eqnarray}
with effects of finite pulse duration $T$ depending on the specific shape of the functions $\kappa_{1}(t),\kappa_2(t)$.

\subsection{Vanishing noise contribution}
We now show that the noise contribution at the end of the QST vanishes due to opposite contributions from $G_1(t_f,t')$ and $G_2(t_f,t')$ in Eq.~\eqref{eq:smdefG}, leading to 
the operator mapping $a_2(t_f) = -a_1(t_i)$.
We first note that the state of the injected noise field can always be decomposed as a distribution of classical states by expressing the injected noise field with a P-representation~\cite{gardiner2004quantum}. If we consider any particular component $\ket{\{\beta(\omega)\}}$ with $b_R(t)\ket{\{\beta(\omega)\}}=\beta(t)\ket{\{\beta(\omega)\}}$, we can bound the noise contribution using the Cauchy-Schwarz inequality: 
\begin{eqnarray}
\left| \int_{t_i}^{t_f}dt'\, G(t_f,t') b_{R}(t')\right|^2 \le B(t_f,t_i) \int_{t_i}^{t_f}dt'\, |G(t_f,t') |^2,  \nonumber
\end{eqnarray}
where $B(t_f,t_i)=\int_{t_i}^{t_f} |\beta(t')|^2 dt'$ is the integrated noise intensity.
Below we provide two examples of coupling functions $\kappa_1(t),\kappa_2(t)$ satisfying Eq.~\eqref{eq:req1}, and we show that the integral of $|G(t_f,t') |^2$ in the last equation vanishes. 

(i) We first consider the functions
 \begin{eqnarray} \kappa_1(t)&=& \kappa_{\max} \frac{e^{\kappa_{\max}t}}{2-e^{\kappa_{\max}t}} \Theta(-t)+\kappa_{\max}\Theta(t),\label{eq:kappaexp}\\ \kappa_2(t)&=&\kappa_1(-t)\end{eqnarray} where we set for convenience $t_f=-t_i=T/2$ and $\kappa_{\max}$ is the maximum decay rate. These are the coupling functions used in our numerical simulations. The integral reads 
\begin{equation}
\int_{t_i}^{t_f}dt'\, |G(t_f,t') |^2 = \frac{2 \left(e^{\kappa_{\max}T/2}-1\right)}{\left(1-2 e^{\kappa_{\max}T/2}\right)^2},
\end{equation}
which vanishes like $e^{-\kappa_{\max}T/2}$ in the limit $\kappa_{\max}T\gg1$. This implies that, provided the integrated noise intensity does not grow exponentially with $T$ (typically the growth is linear), the noise contribution vanishes. Unless stated otherwise, we use $\kappa_{\max}T=20$.

(ii) This cancellation of noise is not restricted to coupling functions satisfying the time-reverse property $\kappa_2(-t)=\kappa_1(t)$. For example if one considers
\begin{eqnarray}
\kappa_1(t) &=& \kappa_{\max} \\
\kappa_2(t) &=& \kappa_{\max} \frac{e^{-\kappa_{\max}t}}{1-e^{-\kappa_{\max}t}},
\end{eqnarray}
where we set $t_i=0$ and $t_f=T$, then the integral reads 
\begin{equation}
\int_{t_i}^{t_f}dt'\, |G(t_f,t') |^2 = e^{-\kappa_{\max}T}.
\end{equation}
Note however that in practice one needs to introduce a cut-off for $\kappa_2(t)$ to avoid the divergence at time $t=0$. 

\subsection{Extension: beamsplitter operation}\label{sec:bsop}
\begin{figure}
\includegraphics[width=\columnwidth]{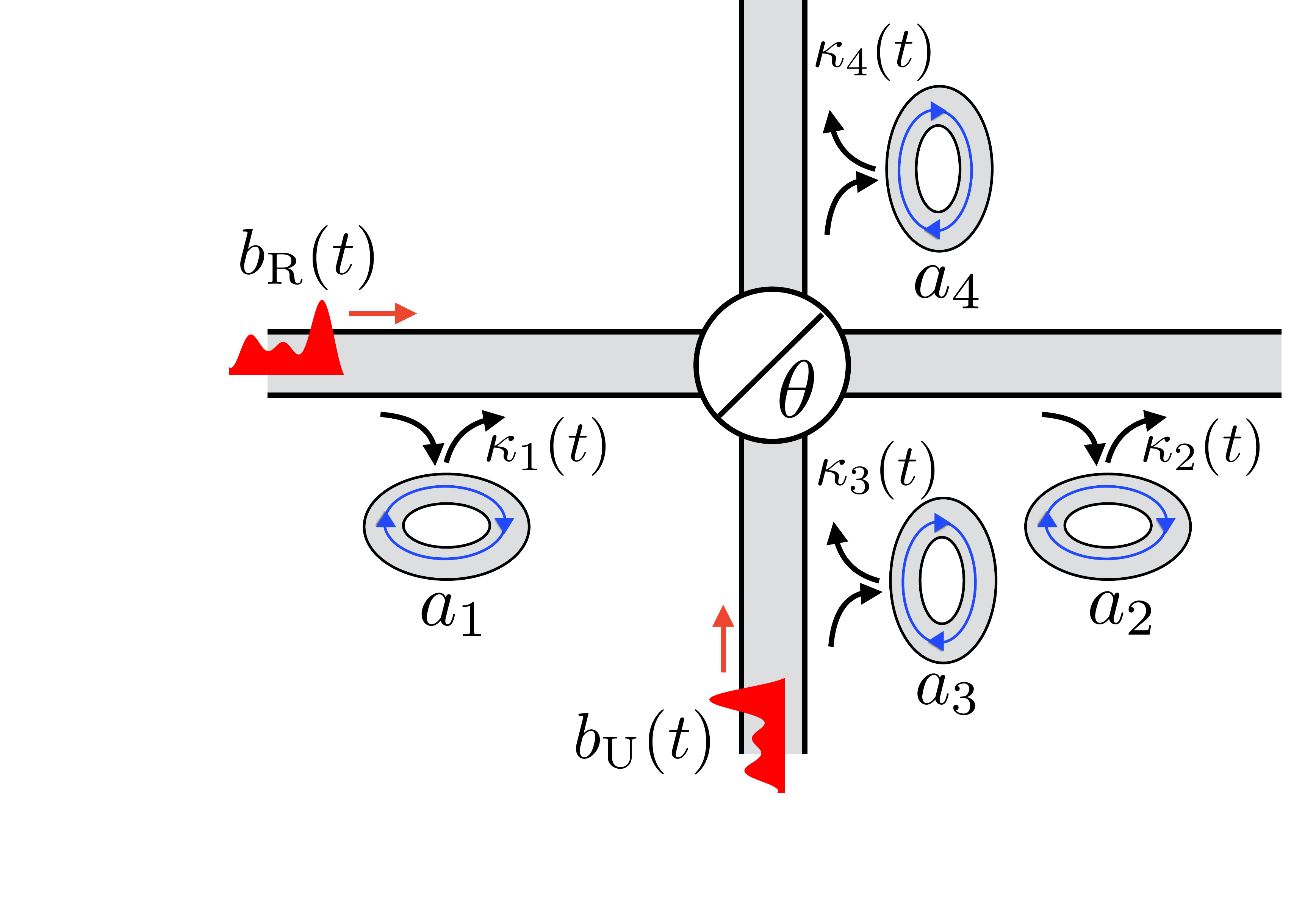}
\caption{{\it Operator mapping between four cavities.} We consider four cavities in the beamsplitter configuration and subject to injected noise. \label{fig:beamsplitter}}
\end{figure}
The operator mapping between two nodes can be extended to the beamsplitter mixing between four nodes, as represented in Fig.~\ref{fig:beamsplitter} where two pairs of cascaded nodes as harmonic oscillators undergo the QST protocol along waveguides mixed by a beamsplitter.

If we assume no imperfections in the couplings, the QLEs read
 \begin{eqnarray} 
 \dot a_1 =&& -\frac{\kappa_1(t)}{2}a_1(t) - \sqrt{\kappa_1(t)}b_{R}(t) \\
\dot a_2 =&& -\frac{\kappa_2(t)}{2}a_2(t) -\sqrt{\kappa_2(t)}\Big(\sqrt{\kappa_1(t)}a_1(t)\cos(\theta)\\&&\nonumber  + b_{R}(t)\cos(\theta) - \sqrt{\kappa_1(t)}a_3(t)\sin(\theta) - b_{U}(t)\sin(\theta)\Big),\\ 
\dot a_3 =&& -\frac{\kappa_3(t)}{2}a_3(t) - \sqrt{\kappa_3(t)}b_{U}(t) \\ \label{eqdam4}
\dot a_4=&& -\frac{\kappa_4(t)}{2}a_4(t) -\sqrt{\kappa_4(t)}\Big(\sqrt{\kappa_3(t)}a_3(t)\cos(\theta)\\&&\nonumber  + b_{U}(t)\cos(\theta) + \sqrt{\kappa_3(t)}a_1(t)\sin(\theta) + b_{R}(t)\sin(\theta)\Big).
   \label{eq:langevinbeamsplitter} 
   \end{eqnarray}
 Here nodes $3$ and $4$ are coupled only to the modes propagating upwards in the vertical waveguide, and $b_{U}(t)$ denotes the corresponding injected noise. In particular, if the coupling functions satisfy $\kappa_3(t)=\kappa_1(t)$ and $\kappa_4(t) = \kappa_2(t)$, the equations can be decoupled using the linear combinations
\begin{eqnarray}\tilde a_2(t) &=& \cos(\theta)a_2(t)+\sin(\theta) a_4(t),\\\tilde a_4(t) &=& \cos(\theta)a_4(t)-\sin(\theta) a_2(t), \end{eqnarray}
whose evolution maps directly to Eq.~(2) of the main text with the set ($a_1,a_2$) replaced by the sets ($a_1,\tilde a_2$) or ($a_3,\tilde a_4$). The operator mapping can thus be applied to these sets and we obtain the mixed QST
\be \begin{pmatrix} a_2(t_f) \\ a_4(t_f) \end{pmatrix} = - \begin{pmatrix} &\cos(\theta) &-\sin(\theta) \\ &\sin(\theta)  &\cos(\theta) \end{pmatrix} \begin{pmatrix} a_1(t_i)\\ a_3(t_i) \end{pmatrix}. \label{eq:beamsplitter}\ee 
This result shows that one can realize beamsplitter mappings between four distant cavities, where cavities $1$ and $3$ imprint their quantum states onto noisy temporal modes of the two quantum channels. These modes interfere at the beamsplitter, similarly to single photon wave-packets, before being reabsorbed perfectly by cavities $2$ and $4$.
 
 \section{Effective tunable cavity couplings $\kappa_{1,2}(t)$ via atomic ensembles}\label{sec:atomicensembles}
\subsection{Model}
\begin{figure}
\includegraphics[width=\columnwidth]{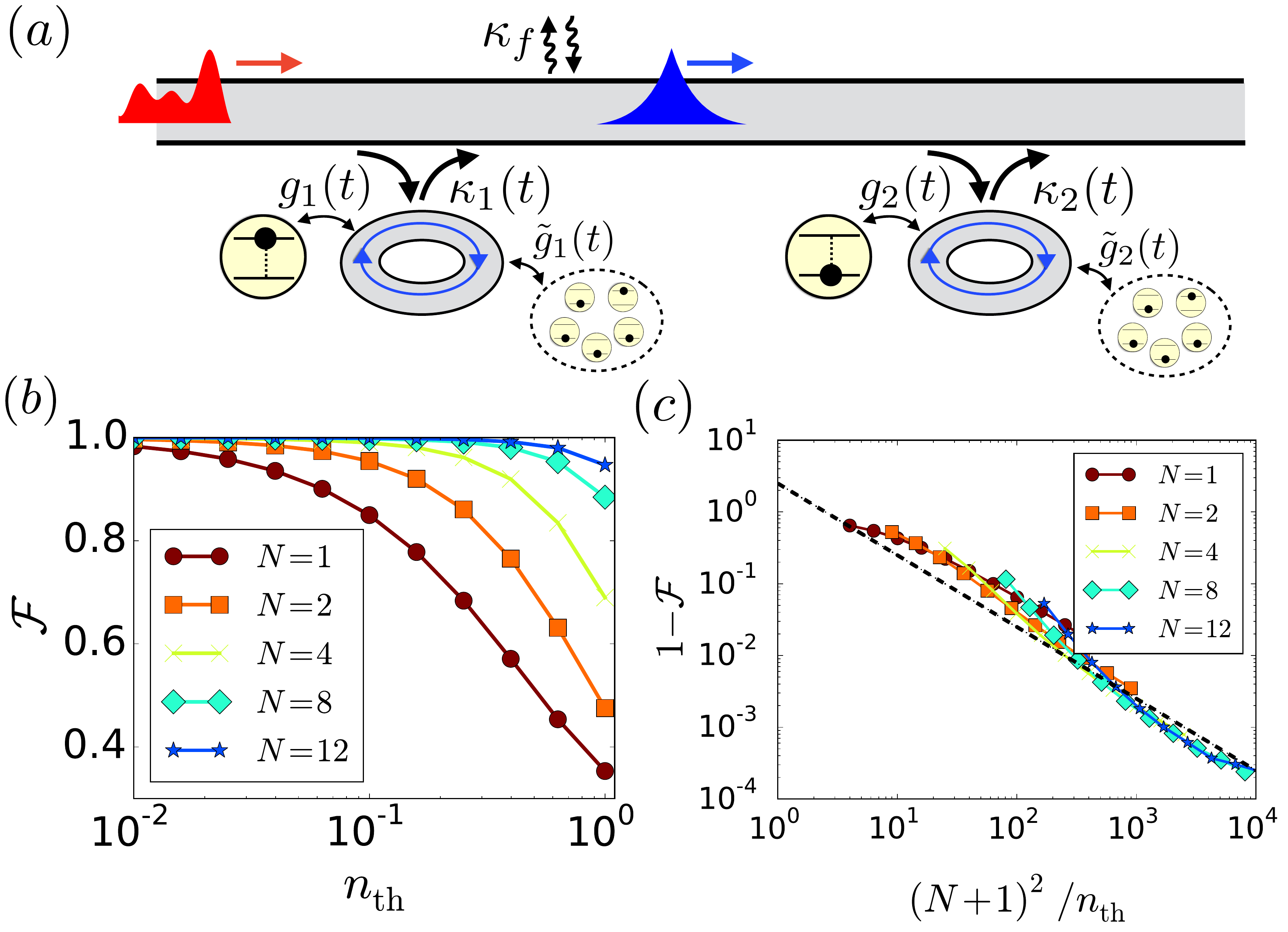}
\caption{{\it QST via atomic ensembles.} (a) Setup including atomic ensembles as effective harmonic oscillators in the nodes. (b) Fidelity of QST as a function of $n_\text{th}$ and for different atom numbers. (c) Error $1-\mathcal{F}$ as a function of the scaling parameter $x=(N+1)^2/n_{th}$. The dashed line represents $2.5/x$.}\label{fig:ensembles}
\end{figure}

Here we provide details on the realization of QST between two nodes ($j=1,2$) using atomic ensembles as harmonic oscillators.
As depicted in Fig.~\ref{fig:ensembles}(a), we consider two nodes consisting of an atomic qubit, a cavity, and an ensemble of two-level atoms $i=1,..,N$ with ground states $\ket{g}_{j,i}$ and excited states $\ket{e}_{j,i}$.
The Hamiltonian governing the dynamics of the system can be written in the form of $H(t)$ (see main text) where the node Hamiltonian is now given by
\begin{equation}
H_{n_j} = g_j \left( a_j^\dagger \sigma_j^- +  \text{H.c} \right)+\tilde g_j(t) \sqrt{N} \left( a_j^\dagger S_j^- +  \text{H.c} \right), 
\end{equation}
and, having in mind a typical quantum optical setup, the coupling rates between cavities and waveguide $\kappa_j$  are considered to be constant over time.
The node Hamiltonian $H_{n_j}$ can be realized via a laser-assisted Raman transition~\cite{Hammerer2010} with the assumption that additional Stark-shifts terms can be cancelled using for instance another laser coupling.
Here $S_j^{\pm}=(S_j^x\pm i S_j^y)/\sqrt{N}$ are the collective spin operators associated to the total angular momentum $S_j^l=\frac12\sum_{i=1}^N \sigma_{i,j}^l$, with spin $S=N/2$ and where $\{\sigma_{i,j}^l,l=x,y,z\}$ denotes the set of Pauli matrices of the $i$-th atom in node $j$. In the following we will use the ground states $\ket{G}_j=\otimes_{i}\ket{g}_{j,i}$ and first excited states $\ket{E}_j=S_j^+\ket{G}_j$ of the atomic ensemble to mediate the QST between the two nodes, with effective time-dependent coupling to the waveguide governed by the functions $\tilde{g}_j(t)$.

The  protocol, which becomes robust against noise in the limit of large atomic ensembles, is very similar to the one described in the main text in Fig.~1(b). 
The first step (i) now consists in mapping the qubit state $c_g\ket{g}_1+c_e\ket{e}_1$  of the atom to a collective state of the atomic ensemble $c_g\ket{G}_1+c_e\ket{E}_1$.
This can be done for instance by detuning the atom and atomic ensemble with respect to the cavity in the Raman transitions, hence realizing an effective coupling between qubit and atomic ensemble with the cavity adiabatically eliminated.
The second step (ii) consists in realizing the operator mapping $S_1^-(t_i)\to -S_2^-(t_f)$ in analogy to the operator mapping between cavities, with the atomic ensemble now resonantly coupled to the cavity. Finally, (iii) the atomic ensemble state is mapped to the qubit state of the second node in the reverse process of step (i).

\subsection{Adiabatic elimination of the cavities}

We now describe in details step (ii) where the atomic ensemble realizes the QST in presence of injected noise. During this operation, the qubits are excluded from the dynamics ($g_j=0$). The Langevin equations associated with $H(t)$ can be written in the form (see for instance Ref.~\cite{Habraken2012} for the single atom case)
\begin{eqnarray}
\dot S^-_j &=&  -i \tilde g_j(t) \sqrt{N}  [S_j^-,S_j^+]  a_j  \nonumber \\
\dot a_j &=& -i \tilde g_j(t) \sqrt{N} S_j^- -\frac{\kappa_j}{2} a_j - \sqrt{\kappa_j(t)} \left(b_{R}(t)+\delta_{j,2} a_1(t) \right), \nonumber
\end{eqnarray}
with $j=1,2$. We assume here the bad-cavity regime where the decay of the cavity $\kappa_j$ to the waveguide is large compared to the coupling $\tilde g_i$. The cavity mode can then be adiabatically eliminated, leading to
\begin{eqnarray}
\dot S_1^- &=& -[S_1^-,S_1^+] \left[\frac{\tilde\kappa_1(t)}{2}   S_1^-+ \sqrt{\tilde \kappa_1(t)} b_{R}(t) \right]\nonumber \\
\dot S_2^- &=& -[S_2^-,S_2^+] \left[\frac{\tilde\kappa_2(t)}{2}  S_2^-  \right. \nonumber \\
&&\hspace{1.6cm}+\left.  \sqrt{\tilde \kappa_2(t)}  \left( b_{R}(t) + \sqrt{\tilde \kappa_1(t)} S_1^- \right) \right], \nonumber \\ \label{eq:langevincollec}
\end{eqnarray}
with the time-dependent coupling $\tilde \kappa_j(t)=4\tilde g_j(t)^2N/\kappa_j$. We have a absorbed a phase $i$ ($-i$) in the definition of $S_1$ ($S_2$, respectively).

In the case of low excitations $n\ll N$, the spin operators behave as linear operators~\cite{Hammerer2010} where \be [S_j^-,S_j^+]\ket{n}_j=-\frac{2}{N}S_j^z\ket{n}_j=\left(1-\frac{2n}{N}\right)\ket{n}_j\approx \ket{n}_j,\ee with $\ket{n}_j\propto (S_j^+)^n\ket{G}_j$. Eqs.~\eqref{eq:langevincollec} thus map to Eqs.~(2) of the main text with the identification $S_j^-=a_j$, and the ensemble behaves as a harmonic oscillator.

\subsection{Nonlinear effects}

In order to assess the importance of nonlinear effects for finite number of atoms, we simulate the QST dynamics by mapping the QLEs~\eqref{eq:langevincollec} to a master equation, which is then integrated numerically.
In Fig.~\ref{fig:ensembles}(b) we show how the QST fidelity approaches unity in the transition from the nonlinear ($N\approx 1$) to the linear regime ($N\gg 1$). In Fig.~\ref{fig:ensembles}(c), we represent the same data in a logarithmic scale and with a rescaled axis $x=(N+1)^2/n_{th}$. This shows that the error scales a power law $1-\mathcal{F}\propto 1/x$ in the limit of large $x$. 



%

\section{Matrix product state approach to solve the QSSE with injected noise }\label{sec:MPS}

Our numerical approach based on matrix product state (MPS) builds on the techniques developed in Ref.~\cite{Pichler2016}. The state of the system of nodes and photons is described as a MPS~\cite{Schollwock2011} and evolved according to the QSSE. This allows one to (i) simulate the non-Markovian dynamics of the system for large retardation times $\tau$ when the system is not purely cascaded, (ii) inject arbitrary noise, and (iii) have access to the entangled state of the electromagnetic field inside the waveguide.

We first provide the procedure for the scenario, where the nodes consist of two purely cascaded cavities, \ie~we assume perfect chirality. In the interaction picture the system evolves according to the QSSE \be\label{eq:smqsse} i\frac{d}{dt}\ket{\Psi(t)}=V(t) \ket{\Psi(t)}, \ee with $V(t)$ defined in Eq.~(1) of the main text.
We now discretize time $t_{k+1} = t_k + \Delta t$ as time-bins with time-step $\Delta t$, with initial time $t_1=t_i$, final time $t_f=t_M$ and $T=t_f-t_i= M \Delta t$. We require $\Delta t$ to be smaller than the timescale of the system dynamics, \ie~$\kappa_{\max} \Delta t \ll 1$. For each time-bin we define the quantum noise increment \be \Delta B_k=\int_{t_k}^{t_{k+1}} dt\,b_R(t), \ee with $b_R(t)$ the Fourier transform operators of $b_R(\omega)$, which obey bosonic commutation relations $[\Delta B_k, \Delta B_{k'}^\dagger]=\Delta t\,\delta_{k,k'}$. The time-bins thus describe bosons with creation operators $\Delta B_k^\dagger/\sqrt{\Delta t}$, and the operator $\Delta B_k^\dagger \Delta B_k/\Delta t$ is interpreted as the total number of photons with label $t\in [t_k,t_{k+1}]$. 
The time evolution can then be viewed as a stroboscopic mapping $\ket{\Psi(t_{p+1})}=U_p \ket{\Psi(t_p)},$ where the unitary operator is given to lowest order in $\Delta t$ by \begin{equation} U_p =\exp\left( \sqrt{\kappa_1(t_p)} \Delta B_p^\dagger a_1 + \sqrt{\kappa_2(t_{p})} \Delta B_{p-l}^\dagger a_2 -\text{H.c.}\right)\label{eq:defUk} \end{equation} with $\tau=l\Delta t$ and the propagation phase has been absorbed in the definition of $a_2$.
The (entangled) state of the system at each time-step is written as \be \label{eq:psitp}\ket{\Psi(t_p)}=\sum_{i_{n1},i_{n2},\{i_k\}} \psi_{i_{n1},i_{n2},\{i_k\}}(t_p)\ket{i_{n1},i_{n2},\{i_k\}}, \ee where $i_{n1},i_{n2}=0,1,...$ label the Fock states of nodes $1$ and $2$, and $\{i_k\}=\{i_1,i_2,...,i_M\}$ label the Fock states of time-bins $k=1,2,...,M$. 

If the injected noise is in a thermal state, the state is no longer pure, and we have the correlations $\langle b(\omega)^\dagger b(\omega') \rangle =n_\text{th} \delta(\omega-\omega')$ with flat spectrum $S(\omega)=n_\text{th}$, or in the time representation $\langle b(t)^\dagger b(t') \rangle = n_\text{th} \delta(t-t')$. This translates for the noise increments into \be \langle \Delta B^\dagger_k \Delta B_{k'}\rangle = n_\text{th} \delta_{k,k'} \Delta t, \ee where $n_\text{th}$ is now interpreted as the average occupation number of each time-bin. This is generated by having each time-bin $k$ initially in the mixed state \be \rho_k = (1-e^{-\beta \omega_c})e^{-\beta \omega_c {\Delta B^\dagger_k \Delta B_k}/{\Delta t}}, \ee where $e^{\beta\omega_c}=1+1/n_\text{th}$. 
In order to perform the simulations using mixed instead of pure states, one usually either averages the evolved states obtained with a set of stochastic trajectories, or simulate the evolution of the density matrices \cite{Bonnes2014}. Here we employ an alternative method, where we purify the state by extending its definition to auxiliary virtual time-bins. The time-bin $k$ now consists of a pair of one `real' and one `auxiliary' parts, which are initially in the pure state  \be\label{eqqsm:psik} \ket{\psi_k}=\sqrt{1-e^{-\beta \omega_c}}\sum_{n=0}^\infty e^{-\beta  \omega_c n/2} \ket{n}_\text{real}\ket{n}_\text{aux}. \ee The original density matrix $\rho_k$ is obtained by tracing out the auxiliary part, which is done at the end of the evolution when computing observables. We thus want to solve the evolution of a state of the form of Eq.~\eqref{eq:psitp}, where now the photon indices $i_k$ are multi-indices \mbox{$i_k=(i_k^r,i_k^a),$} where $i_k^r$ and $i_k^a$ are indices on the Fock spaces of the real and auxiliary parts of time-bin $k$.

The MPS ansatz consists in writing each coefficient of $\psi_{i_{n1},i_{n2},\{i_k\}}(t_p)$ as the trace of a product of $M+2$ tensors, such as \be\label{eq:defmps} \psi_{i_{n1},i_{n2},\{i_k\}}=\text{Tr}\left(A[n_1]^{i_{n1}} A[n_2]^{i_{n2}} A[1]^{i_1} ...\, A[M]^{i_M} \right). \ee
where each object $A[\cdot]$ is a tensor. As long as the entanglement in the system is low enough, the matrices (bond) dimensions, which are changed dynamically during the evolution, can be truncated by a value $D_{\max}$, which restricts the spanned Hilbert space. 

The order of the matrices [\cf~Eq.~\eqref{eq:defmps}] is arbitrary, that is, one can always find an MPS decomposition with matrices in a given order. For each time $t_p$ we chose the following ordering. The time-bin matrices are ordered in order of increasing indices. The matrix for node $1$ is located on the left of $A[p]^{i_{p}}$, and for node $2$ on the left of $A[p-l]^{i_{p-l}}$. The unitary evolution for time-step $p$ [\cf~Eq.~\eqref{eq:defUk}] thus requires to update only these four matrices. The initial thermal state of the matrices for the time-bins $A[k]^{i_k}$ can be decomposed as a product of matrices for real and auxiliary parts as \be A[k]^{i_k}=B[k]^{i_k^r}\Lambda[k]C[k]^{i_k^a}, \ee where \be(B[k]^{i_k^r})_{i,j}=\delta_{i,0}\delta_{j,i_k^r},\ \ \ (C[k]^{i_k^a})_{i,j}=\delta_{j,0}\delta_{i,i_k^a},\ee and \be(\Lambda[k])_{i,j}=\delta_{i,j}\sqrt{1-e^{-\beta\omega_c}}e^{-\beta \omega_c i/2}\ee contains the Schmidt values of the singular value decomposition of Eq.~\eqref{eqqsm:psik}. The algorithm then consists in updating the tensors $B[k],A[n_1],A[n_2]$ by successively applying operators $U_p$.

This model can immediately be extended to more elaborate nodes, e.g. adding atoms coupled to the cavities, by adding other contributions to the Hamiltonian in the QSSE~\eqref{eq:smqsse} and adapting accordingly the Hilbert spaces of the node and the definition of $U_p$ in Eq.~\eqref{eq:defUk}. In practice we obtain the fidelity of QST as described in Sec.~\ref{sec:avfid}, by adding an additional node, decoupled from the dynamics, which is initially in a maximally entangled state with node $1$.

If the nodes couple also to photons propagating in the left direction, one needs to distinguish between the left- and right-moving photons. This evolution is generated by a unitary operator which now reads \cite{Pichler2016}
\begin{equation} \begin{aligned}U_p =\exp\Big(& \sqrt{\beta\kappa_1(t_p)} \Delta B_{\text{R},p}^\dagger a_1 + \sqrt{\beta\kappa_2(t_p)} \Delta B_{\text{R},p-l}^\dagger a_2 \\
+&\sqrt{(1-\beta)\kappa_1(t_p)} \Delta B_{\text{L},p-l}^\dagger a_1 \\+& \sqrt{(1-\beta)\kappa_2(t_p)} \Delta B_{\text{L},p}^\dagger a_2 e^{-2i\phi}  -\text{H.c.}\Big),\label{eq:defUk_notcascade} \end{aligned} \ee
where we define noise increments for right- and left-moving photons as $\Delta B_{\text{R/L},k}=\int_{t_k}^{t_{k+1}}dt\,b_\text{R/L}(t)$. Now the definition of the system extends to the time-bins for left-moving photons. A detailed description of the ordering of the matrices in the MPS decomposition, as well as the treatment of the long-range interactions induced by the time-delay can be found in the Supplemental Material of Ref.~\cite{Pichler2016}. The same purification technique can be applied for both types of photons in order to represent thermal noise.

\section{Frequency mismatch}\label{sec:mismatch}
We study here the effect of a mismatch in the frequencies of the cavity modes, assuming that they are detuned with a frequency $2\Delta$. In the numerical simulations this is described by an additional Hamiltonian term $\Delta(a^\dagger_1 a_1 - a^\dagger_2 a_2)$. In the frame rotating with this new term, we replace $\sqrt{\kappa_2(t)}\to\sqrt{\kappa_2(t)}e^{-2i\Delta t}$ in the QLEs (Eq.~(2) of the main text). The mismatch thus acts as an error on the coupling functions, hence one would now need to add a phase in the coupling between cavities and waveguide in order to emit and absorb the photon in the correct temporal mode.

In Fig.~\ref{fig:mismatch}(a) we see that the effect is similar to the one of timing errors represented in Fig.~2(b) of the main text. Here we need $\Delta/\kappa_\text{max}\lesssim 0.025$ in order to ensure a fidelity of $\mathcal F\geq0.99$. For comparison we show the same plot in the case where the qubits are coupled directly to the waveguide, with $\Delta$ now the detuning between the qubits.
\begin{figure}
\includegraphics[width=\columnwidth]{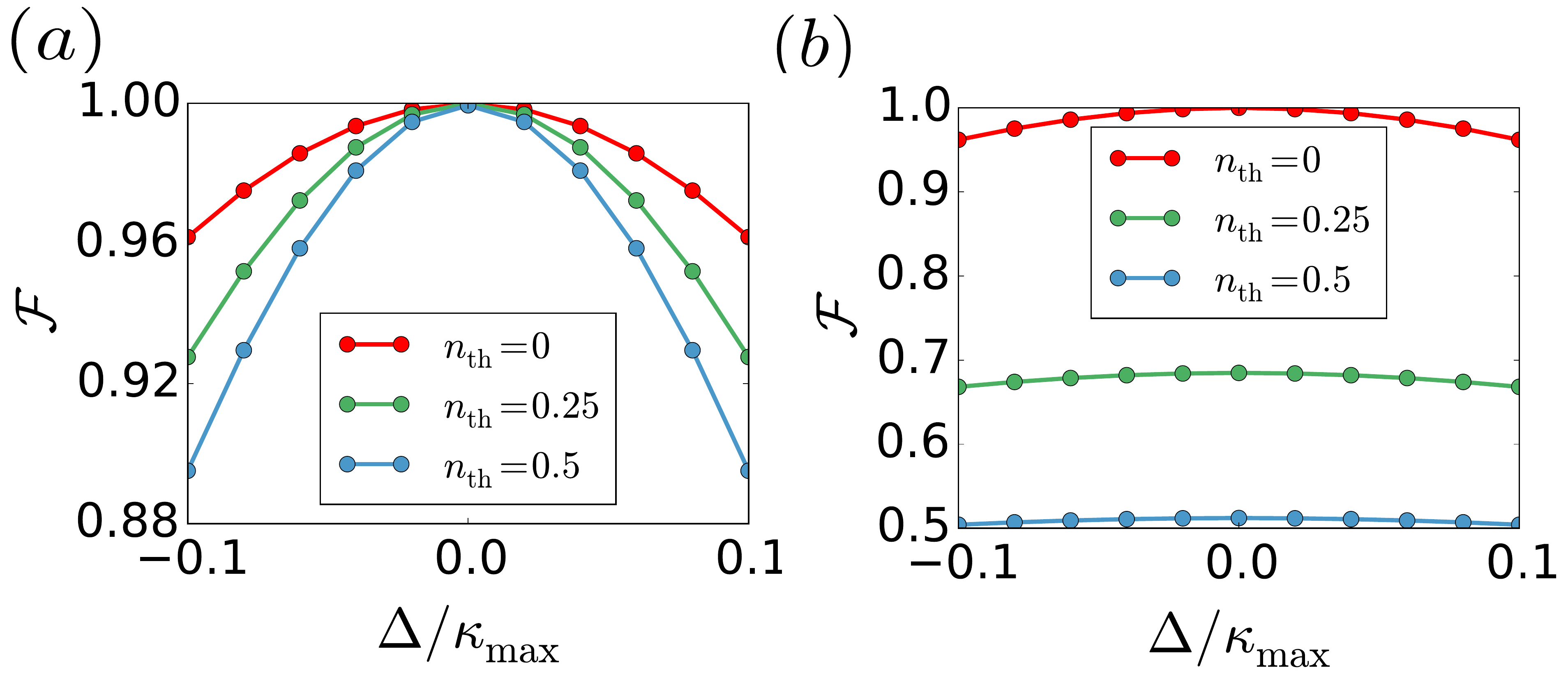}
\caption{{\it Effect of frequency mismatch.} Average fidelity of the state transfer as a function of the detuning $\Delta$ for different values of $n_{th}$, (a) with and (b) without cavities [see respectively Fig.~1(b) and Fig.~1(a) of the main text].}\label{fig:mismatch}
\end{figure}

\section{Quantum error correction}\label{sec:qec}
Here we provide details on the quantum error correction scheme. 
\subsection{Operator mapping with waveguide losses}\label{sec:oplosses}
We first consider the case of waveguide losses, with rates $\kappa_f$. 
These losses can be modelled as a beamsplitter operation, as presented in Sec.~\ref{sec:bsop}, with the angle defined as $\cos(\theta)=\exp(-\kappa_f\tau/2)$. Here the operators $a_3$ and $ a_4$ represent fictitious cavities. Using the beamsplitter mapping of Eq.~\eqref{eq:beamsplitter}, an initial state of the first cavity $\ket{\Psi}_i = \sum_{n=0}^\infty f_n (a_1^\dagger)^n\ket{0}$ is mapped via the QST protocol to the state
\be \label{eq:psifdef}\ket{\Psi_f} = \sum_{n=0}^\infty f_n (-a_2^\dagger \cos(\theta) -a_4^\dagger \sin(\theta))^n \ket{0}, \ee 
where the term associated to $a_4$ represents the photons lost in the process. 
By tracing over the state of the fictitious cavity $a_4$, we obtain the density matrix
\begin{equation}
\rho_f = \rho_0 + \rho_{-1}+ \rho'_f
\end{equation}
of the second cavity with $\rho_0 = \ket{\Psi_0}\bra{\Psi_0}$, \mbox{$\rho_{-1} = \mathcal{P}\ket{\Psi_{-1}}\bra{\Psi_{-1}}$} and $\rho'_f$, corresponding respectively to the cases of no photon loss, one loss and more than one loss. Here, $\ket{\Psi_0} = \sum_n  f_n (-\cos(\theta) a_2^\dagger)^n \ket{0}$,  $\ket{\Psi_{-1}} =a_2 \ket{\Psi}_0$ and \mbox{$\mathcal{P}=\sin^2(\theta)=1-\exp(-\kappa_f\tau)$} is the single-photon loss probability.
Remarkably, the result is still independent of the injected noise.

\subsection{Application to quantum error correction}

We now provide additional details on the simulation of the QST protocol including QEC. 
In the case of a code protecting against single photon losses [\cf~Fig.~2(e) of the main text], the state of the first cavity is written in the orthogonal  basis $\ket{\pm}_1=(\ket{0}_1\pm \sqrt{2}\ket{2}_1+\ket{4}_1)/2$.
After the transfer, the error is detected by realizing a projective measurement of the photon parity number in the second cavity.
We then apply a unitary $U_{0}$ or $U_{-1}$ depending on the measurement outcome $p=0,-1$ [for even and odd respectively], which maps the state of the cavity to the qubit as
\begin{eqnarray}
\Big(c_g\! \ket{-^{(p)}}_2\!+\! c_e \ket{+^{(p)}}_2\Big) &\otimes& \ket{g}_2 \nonumber \\ &\downarrow&  \\ \ket{-^{(p)}}_2 &\otimes& \Big(c_g\! \ket{g}_2\! + \!c_e \ket{e}_2\Big) \nonumber.  \label{eq:unitary} 
\end{eqnarray}
Here $\ket{\pm^{(0)}}=\ket{0}\pm\sqrt{2}\cos^2(\theta)\ket{2}+\cos^4(\theta)\ket{4}$ and $\ket{\pm^{(-1)}}=a_2\ket{\pm^{(0)}}$, up to normalization factors.
The factors $\cos^n\theta$ arise from the deterministic decay of the Fock state components.
Note that for $\theta\neq0$, the states $\ket{\pm^{(p)}}$ are no exactly orthogonal so that the unitary operation realizes the process of Eq.~\eqref{eq:unitary} only approximately. 
As shown in Fig.~2(e) of the main text, the QEC approach applies to cavity losses.

Finally in the case of external photon additions [\cf~Fig.~2(f) of the main text], we can adapt the above procedure using instead the basis \mbox{$\ket{\pm}_1=(\ket{0}_1\pm\sqrt{2}\ket{3}_1+\ket{6}_1)/2$} to encode the state of the first cavity and measuring the photon number modulo~$3$.
For the measurement outcome $p=0$ (modulo 3), the basis states of the second cavity are \mbox{$\ket{\pm^{({0})}}_2=(\ket{0}_2\mp\sqrt{2}\cos^3(\theta)\ket{3}_2+\cos^6(\theta)\ket{6}_2)$} up to normalization factors. 
For $p=-1$, the basis states are $\ket{\pm^{({-1})}}_2=a_2\ket{\pm^{({0})}}_2$, corresponding to a single photon loss. 
Finally, for \mbox{$p=1$}, we obtain $\ket{\pm^{({1})}}_2=a_2^\dagger \ket{\pm^{({0})}}_2$ meaning a photon is added.

\subsection{Efficiency with coherent cat states}

Compared to the binomial codes given above, `cat codes' are based on non-orthogonal states for $\ket{\pm}_1$, where the first cavity state is encoded as superposition of coherent states, or cat states: \begin{equation*}\ket{\pm}_1=(\ket{\alpha_\pm}_1+\ket{-\alpha_\pm}_1)/\mathcal{N}\mathcal{\pm},\end{equation*} with $\alpha_+= \alpha$, $\alpha_-=i\alpha$ and $\mathcal{N}_\pm$ is a normalization factor. These states can however be realized in current experimental setups by projective measurement of the photon number parity~\cite{Haroche2006,Ofek2016}.
In Fig.~\ref{fig:cat}, we compare the fidelity of the cat states encoding with the binomial code for waveguide losses. The fidelity of the cat code is always lower than the one of the binomial code. 
However, we note that, compared with the absence of correction (black solid line) and for the optimal value of $\alpha$ (here $\sqrt{2}$ as blue dashed line), this code is efficient for loss probabilities up to $\mathcal{P}\approx 0.25$.

\begin{figure}
\includegraphics[width=0.8\columnwidth]{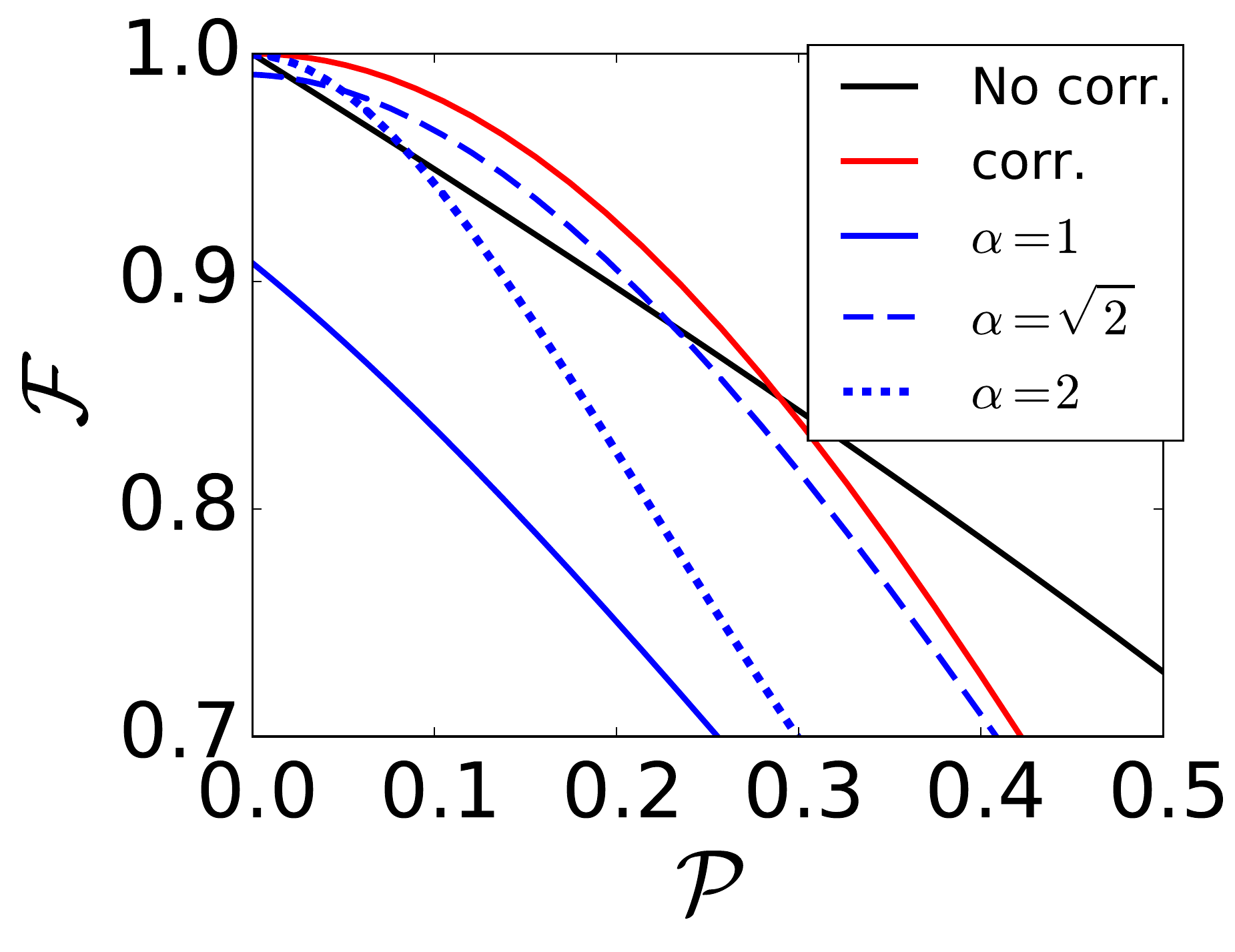}
\caption{{\it Quantum error correction with cat states in the QST protocol.} (a) Fidelity as a function of the loss probability $\mathcal{P}$ considering exclusively waveguide losses ($\kappa'=0$). The non-corrected code is shown in black, the binomial code in red and the cat code in blue with the solid, dashed, dotted lines corresponding respectively to $\alpha=1,\sqrt{2},2$. \label{fig:cat}}
\end{figure}

\section{Quantum state transfer in closed systems}\label{sec:closed}
We describe the model corresponding to the setup depicted in Fig.~3(a) of the main text, where two cavities are coupled to the extremities of a waveguide of length $L$ with discretized modes $b_n$. Thus our setup consists of a multimode cavity sandwiched between two single mode cavities.
The Hamiltonian is given by~\cite{Pellizzari1997}
\begin{eqnarray}
H_\mathrm{closed} &=& \sum_{j=1,2}\left(g_j(t) a_j\sum_n  (-1)^{(j-1)n} b^\dagger_n+\textrm{H.c.}\right) \nonumber \\
&+&  \sum_n  n \delta  b^\dagger_n  b_n, \label{eq:Hfiber}
\end{eqnarray}
with the mode spacing $\delta = \pi c/L$ and where the waveguide mode $n=0$ is resonant with the two cavity modes $ a_{1,2}$. In the Heisenberg picture, the formal solution for $b_n$ reads
\begin{eqnarray}
b_n &&= b_n(t_i)e^{-i \delta n (t-t_i)} \nonumber \\ 
&&- i \int_{t_i}^t dt' e^{-i \delta n (t-t')} \left( g_1(t')a_1(t')+(-1)^n g_2(t') a_2(t') \right). \nonumber
\end{eqnarray}
Based on this solution, the equation of motion of $a_j$ reads
 \begin{eqnarray}
 \dot a_j &&= -i g_j(t) \sum_n b_n(t_i) e^{-i \delta n (t-t_i)}  \nonumber \\
 && -\frac{2 \pi g_j(t)}{\delta}\sum_{k=0}^\infty c_k g_j(t-2k\tau )a_j(t-2k\tau) \nonumber \\
 &&  -\frac{2 \pi g_j(t)}{\delta}\sum_{k=0}^\infty  g_{2-j}(t-(2k+1)\tau) a_{2-j}(t-(2k+1)\tau),  \nonumber \\ \label{eq:langevinclosed}
 \end{eqnarray}
where $c_0=1/2$, $c_{k>0}=1$, and the time delay is \mbox{$\tau=\pi/\delta=L/c$}. We used  the Fourier series of the Dirac comb function \mbox{$\sum_n e^{2\pi i n t/T}=T \sum_{k=0}^\infty \delta(t-kT)$}.

In order to obtain the function $g_j(t)$ associated with the QST pulses $\kappa_j(t)$, we first consider a situation where the second cavity and the waveguide are initially in their vacuum state $\big(a_2\ket{\Psi(t_i)}=b_n\ket{\Psi(t_i)}=0\big)$. For times $t_i<t<t_i+2\tau$, Eq.~\eqref{eq:langevinclosed} is then equivalent to Eq.~(2) of the main text with
\begin{equation}
\kappa_j(t)=2\pi g_j^2(t)/\delta. \label{eq:kappaclosed}
\end{equation}
We can now test the robustness of the QST protocol against noise, which we simulate in this closed system by considering each mode of the waveguide to be initially in a coherent state. 
With respect to the definitions of App.~\ref{sec:avfid}, the initial state reads
\begin{equation}
\ket{\Psi(t_i)}=\frac{(\ket{0}_1\ket{0}_a+ \ket{1}_1\ket{1})_a \otimes_n \ket{\alpha}_n \otimes \ket{0}_2}{\sqrt{2}}, 
\end{equation}
where $\ket{\alpha}_n$ is a coherent state for the waveguide mode $n$. Finally, in order to study the effect of cavity non-linearities, we include in $H_\mathrm{closed}$ a Kerr nonlinearity described by the term $-\chi_{j=1,2} \sum_j a_j^\dagger  a_j^\dagger a_j a_j$. We then solve the Hamiltonian dynamics, with $g_j(t)$ varying according to Eq.~\eqref{eq:kappaclosed} and the time-symmetric functions $\kappa_j(t)$. The simulations of this system are in general numerically demanding. However, considering the single mode limit $\delta\gtrsim\kappa_{\max}$, only a few modes of the waveguide are excited. In our case, we chose $\kappa_{\max}=0.3\,\delta$, which allows us to restrict the simulation to only three modes $n=-1,0,1$.

%

\end{document}